\documentclass[%
   reprint,
 amsmath,
 amssymb,
 aps,
prb,
]{revtex4-1}

\usepackage{color}	
\usepackage{graphicx}
\usepackage{dcolumn}
\usepackage{bm}
\usepackage{braket}

\newcommand{\Tsc}{\ensuremath{T_{\rm sc}}}
\newcommand{\TD}{\ensuremath{\Theta_{\rm D}}}
\newcommand{\kB}{\ensuremath{k_{\rm B}}}

\newcommand{\TaTe}{Cu-doped Ta$_{1.6}$Te}

\newcommand{\rev}[1]{#1}

\newcommand{\II}{I\hspace{-.1em}I}

\begin{document}

\preprint{APS/123-QED}

\title{Thermodynamic evidence for full-gap superconductivity in the dodecagonal quasicrystal {\TaTe}}

\author{N. Kabeya$^{1}$}
 \email{kabeya\_fgel@cs.u-ryukyu.ac.jp}
\author{Y. Tokumoto$^{2}$}
\author{K. Tomiyama$^{2}$}
\author{N. Kimura$^{3}$}
\author{K. Edagawa$^{2}$}

\affiliation{$^1$Department of Physics and Earth Science, University of the Ryukyus, Okinawa 903-0213, Japan}
\affiliation{$^2$Institute of Industrial Science, The University of Tokyo, Tokyo 153-8505, Japan.}
\affiliation{$^3$Department of Physics, Tohoku University, Sendai 980-8578, Japan}





\date{\today}
\begin{abstract}
  We report the superconducting gap in the van der Waals layered quasicrystal {\TaTe}, using a fast relaxation technique that removes the large nuclear contribution of $^{181}$Ta.
  The initial-slope method enabled detection of the electronic specific heat down to 60~mK, revealing a fully gapped state with $\Delta(0)/k_{\rm B} = 1.43$~K.
  Both the gap magnitude and specific-heat jump are smaller than BCS predictions, while quasiparticle excitations are strongly suppressed, consistent with a theoretical expectation for aperiodic systems.
  AC-susceptibility measurements show a large upper critical field and pronounced anisotropy, reflecting the quasi-two-dimensional structure.
  These results provide the first thermodynamic evidence for a full-gap superconducting state in a quasicrystal and highlight unconventional pairing mechanisms beyond periodic lattices.


\end{abstract}

\pacs{Valid PACS appear here}
\maketitle

  Superconductivity (SC) is ubiquitously observed in various crystals as well as in amorphous solids\cite{Webb2015, Scanlan2004, Stewart2011, Qiu2021, Rahman2015, Johnson1979}.
  Recent studies have revealed SC in quasicrystals\cite{Kamiya2018, Tokumoto2024} which are characterized by aperiodic long-range order of atoms and forbidden rotation symmetries absent in periodic lattices\cite{Shechtman1984, IUC1992}.
  This discovery demonstrates not only the formation of Cooper pairs under aperiodic potentials but also the emergence of long-range order in the aperiodic electronic eigenstates.

  In contrast to periodic systems, the mature of the electronic eigenstates in aperiodic systems remains under debate.
  They are generally considered to be critical states, intermediate between itinerant and localized states\cite{Kohmoto1987, Fujiwara1989}, whereas their counterparts in periodic systems are the well-established Bloch states defined in reciprocal space ($k$-space).
  Consequently, the Fermi surface in aperiodic systems is not defined in $k$-space, and the Cooper pairs in quasicrystals are unlikely to be described by conventional ${\pm \bf k}$ pairing\cite{Anderson1959}.
  Considering these distinct characteristics, theoretical investigations have been actively pursued, and a number of intriguing superconducting properties specific to quasicrystals have been proposed
  \cite{Fulga2016, Sakai2017, Sakai2019, Takemori2020, Cao2020, Ghadimi2021, Fukushima2023}.

  The defining feature of SC is its order parameter, the superconducting gap, which develops below the transition temperature {\Tsc} and governs quasiparticle excitations at low temperatures.
  In periodic systems, the gap symmetry directly reflects the pairing symmetry of Cooper pairs: the $s$-wave pair yields an isotropic gap in $k$-space, whereas $p$- or $d$-wave pairs generate anisotropic nodes.
  Identifying the gap symmetry is thus essential for characterizing SC.
  In contrast, in aperiodic systems the gap has so fer been inferred only from the specific-heat jump at {\Tsc}, with no direct evidence for its symmetry.

  The superconducting gap can, in principle, be probed by specific-heat measurements well below {\Tsc}, where the contribution of the quasiparticle excitations beyond the gap are expected to dominant\cite{Tinkham1975}.
  In practice, however, such low-temperature measurements are often obscured by nuclear contributions.
  These contributions, familiar in finite magnetic fields, also appear at zero field due to nuclear quadrupole moments in non-cubic environments.
  Elements such as Lu, Ir, Re, and Ta posses particularly large quadrupole moments, as exemplified in {\TaTe}, as discussed below.


  Ta$_{1.6}$Te is a recently discovered superconducting quasicrystal (QC) with a transition temperature ($\Tsc = 0.98$~K)\cite{Tokumoto2024} higher than that of the Al-Zn-Mg QC ($\Tsc \approx 50$~ mK)\cite{Kamiya2018}.
  This compound forms a dodecagonal aperiodic structure in two-dimensional layers periodically stacked via van der Waals gaps\cite{Conrad1998}.
  Such a structure combines aperiodicity with periodicity, providing a unique platform to explore the differences between  periodic and aperiodic effects on SC.
  In addition, transition-metal dichalcogenides are known for unusual superconducting properties, including two-dimensionality and broken time-reversal symmetry\cite{Ribak2020}.

  %

  In this study, we adopted a recently developed technique for low-temperature specific heat that effectively eliminates the nuclear contribution.
  With this approach, we observed the superconducting gap down to 60~mK ($T/\Tsc = 0.07)$, and confirmed that Ta$_{1.6}$Te dodecagonal QC hosts a fully gaped superconducting state.

  Polycrystalline samples of the {\TaTe} ((Ta$_{0.95}$Cu$_{0.05}$)$_{1.6}$Te) were synthesized using a procedure similar to that for pure Ta$_{1.6}$Te\cite{Tokumoto2024}.
  A small amount of Cu was introduced to suppress contamination phases reported in undoped Ta$_{1.6}$Te, presumably by stabilizing the aperiodic structure.
  Samples used for measurements were taken from the same pellet and formed into a thin plate (\rev{thick of 0.2~mm, area of 15~mm$^2$}) for specific-heat measurements, and a grain (diameter of 1~mm, length of 3~mm) for susceptibility measurements, using emery paper.

  The ac magnetic susceptibility was measured using a mutual induction method with a second-derivative coil as the pick-up.
  The oscillation field was set to 179~Hz and 0.1~mT.
  The magnitude of the susceptibility was calibrated using a lead specimen shaped similarly to the {\TaTe} sample.

  Specific-heat measurements were performed using two techniques:
  the conventional heat pulse method (HPM) and the initial slope method (ISM) we developed recently\cite{ISM2025}.
  Both employed the same calorimetric setup, consisting of a laboratory-made quasi-adiabatic cell.
  The ISM is a relaxation method utilizing only the first few sub-seconds of the relaxation process, based on the technique of Kihara {\it et al.}, following the concept original proposed by Andraka and Takano\cite{Andraka1996, Andraka2011, Kihara2014}.
  With several modifications, we adapted the ISM for dilution-refrigerator temperatures, where minimal heating is essential.
  Further details of the ISM are reported elsewhere\cite{ISM2025}.


  \begin{figure}[bt]
    \includegraphics[width=7cm]{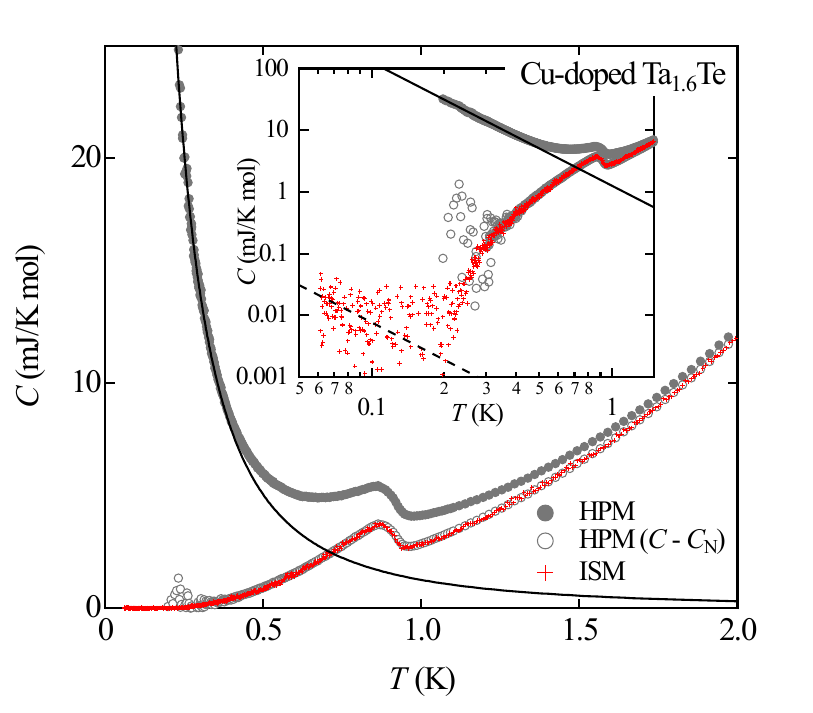}
    \caption{
      Specific heat of {\TaTe} measured by the conventional heat-pulse method (HPM) and the initial-slope method (ISM).
      The solid curve represents an nuclear contribution fitted as $C_{\rm N} \propto T^{-2}$.
      Open circles denote the electronic and phonon contributions obtained by subtracting $C_{\rm N}$.
      The inset shows the same data on double logarithmic axes.
      The additional broken line indicates a nuclear contribution $C_{N}' \propto T^{-2}$ possibly remaining in the results of ISM.
      }
    \label{Fig_HPM_vs_ISM}
  \end{figure}

  Figure~\ref{Fig_HPM_vs_ISM} shows the temperature dependence of the specific heat of {\TaTe} obtained by the HPM and ISM.
  The filled circles represent the HPM data, which diverge at low temperatures due to the nuclear contribution.
  This divergence is well reproduced by a fit $C_{\rm N} = A T^{-2}$, as represented by the solid-curve.
  Subtracting this fitted $C_{\rm N}$ yields the electronic and phonon contributions, $C - C_{\rm N}$, plotted as open circles.
  $C - C_{\rm N}$ decreases monotonically below {\Tsc}, as expected, but scatter strongly below $T \lessapprox 0.4$~K because of subtraction between large values.
  This scattering clearly recognized in the inset, which shows a double-logarithmic plot of the same data.
  By contrast, the ISM results shown by cross markers agree closely with $C - C_{\rm N}$ data without requiring subtraction, since the ISM inherently excludes the nuclear contribution and captures only electronic and phonon contributions.
  Above $\sim 0.4$~K, the ISM and  $C - C_{\rm N}$ data coincide, confirming the accuracy of the evaluation.
  Residual nuclear contributions may remain in the ISM, as suggested by the broken line in the inset of Fig.~\ref{Fig_HPM_vs_ISM}, but they are more than $10^{4}$ times smaller than in the HPM.
  Thus, the ISM provides reliable access to quasiparticle excitations at very low temperatures, enabling our analysis of superconducting gap symmetry.


  \begin{figure}[bt]
    \includegraphics[width=7.5cm]{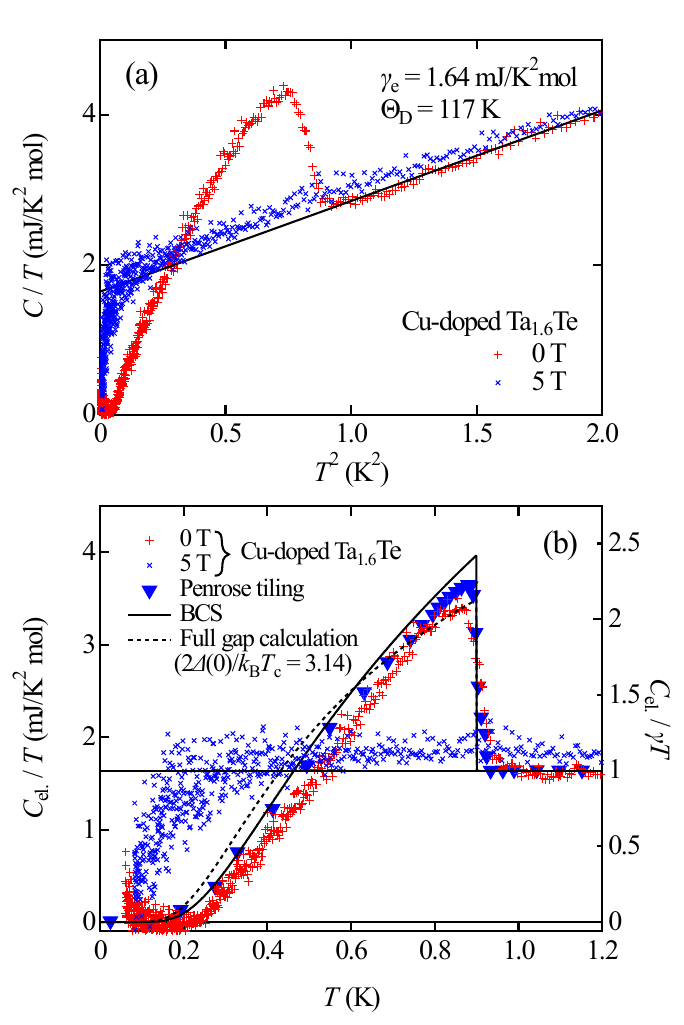}
    \caption{
      (a) $C/T$ vs $T^2$ for {\TaTe} below $T \lessapprox 1.7$~K.
      The solid line is a linear fit between $1.0$ and $1.5$~K used to estimate the phonon and electron contributions in the normal state excluding the contamination phases.
      (b) Electron contribution $C_{\rm el.} / T$ in the superconducting (0~T) and the normal (5~T) states.
      The horizontal line indicates an expected electronic contribution in the normal state.
    }
    \label{Fig_CiT}
  \end{figure}

  We now focus on the physical properties of {\TaTe}.
  First, we evaluate phonon and electronic contributions from the $C/T$ vs $T^{2}$ plot.
  As shown in Fig.~\ref{Fig_CiT}(a), a linear fit yields $\gamma = 1.64$~mJK$^{-2}$mol$^{-1}$ and $\TD = 117$~K.
  Here, the ``mol" unit refers to the number of atoms irrespective of the elements, so one mole of {\TaTe} corresponds to 2.6~mol.
  After subtracting the phonon contribution, the electronic specific heat $C_{\rm el.} / T$ is shown in Fig.~\ref{Fig_CiT}(b).
  The right axis represents $C_{\rm el.} / (T \gamma)$.
  $C_{\rm el.} / T$ falls to zero near 0.25~K and remains there at low temperature, indicating a finite excitation gap and negligible not-superconducting (normal) fractions.
  Furthermore, this result indicate that the small jump height of $\Delta C_{\rm el.} / (\Tsc \gamma) \approx 1.1$ cannot be ascribed to trivial issues, such as contamination phases, but instead reflects an intrinsic property of superconducting phase in {\TaTe}.

  Figure~\ref{Fig_GapFit} shows a semi-logarithmic plot of the specific heat below {\Tsc}.
  As expected from Fig.~\ref{Fig_CiT}(b), specific heat below {\Tsc} follows an exponential behavior.
  The solid line represents a fit for $C \propto \exp \left( - {\Delta(0)} / ({\kB T}) \right)$, which reproduces the experimental result well below $0.5$~K.
  This result clearly suggests a fully gapped superconducting state realizing in {\TaTe}.
  From the slope of the fit, we estimate the gap energy at $T = 0$~K as $\Delta(0)/\kB = 1.43$~K.

  We also examined the entropy balance between the superconducting and normal states.
  Assuming $C_{\rm el.} / T = 0$ below the lowest accessible temperature, the entropy of the superconducting state at 1.0 K is estimated from Fig.~\ref{Fig_CiT}(b) as $S_{\rm SC}(1\,{\rm K}) = 1.45$~mJK$^{-1}$mol$^{-1}$.
  In contrast, the nomal-state entropy, obtained as the horizontal line in Fig.~\ref{Fig_CiT}(b), is $S_{\rm N} = 1.64$~mJK$^{-1}$mol$^{-1}$, larger than $S_{\rm SC}$.
  According to the third law of thermodynamics, these values should be identical.
  The origin of this discrepancy is unclear, but the electronic contribution at 5~T ($>H_{\rm c2}$) also exhibits an unexpected drop below 0.2~K.
  Note that the slightly larger $C_{\rm el.} / T$ under 5~T compared with that under 0~T is attributed to a contribution of a contamination phase with {$\Tsc \sim 3$~K and $H_{\rm c2} < 5$~T.
  The anomalous behavior of $C_{\rm el.} / T$ under 5~T may reflects a fine structure of the pseudogap, as reported in some quasicrystals\cite{Stadnik1997, Tamura2004}, and warrants further investigation.


  \begin{figure}[t]
    \includegraphics[width=7cm]{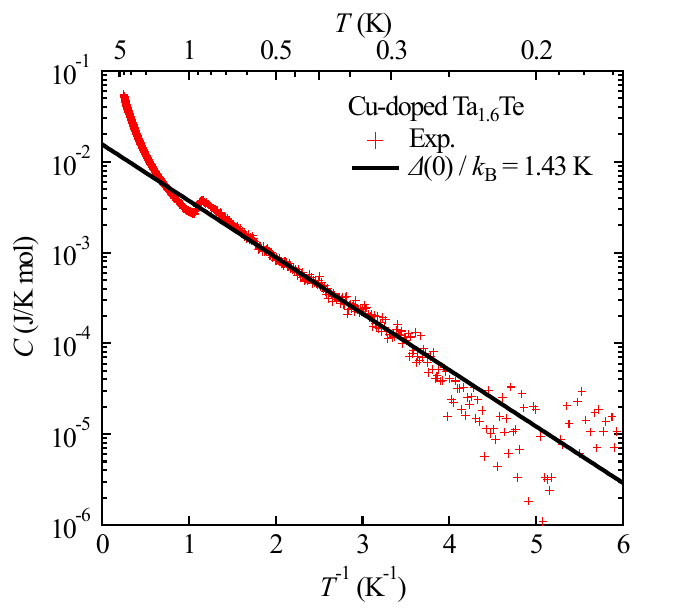}
    \caption{
      Semi-logarithmic plot of the specific heat $C$ vs inverse temperature $T^{-1}$ for {\TaTe}.
      The solid line represents a fit using $C \propto \exp(-\Delta(0)/k_{\rm B}T)$, resulting $\Delta(0)/k_{\rm B} = 1.43$~K.
      }
    \label{Fig_GapFit}
  \end{figure}



  \begin{figure}[t]
    \includegraphics[width=7cm]{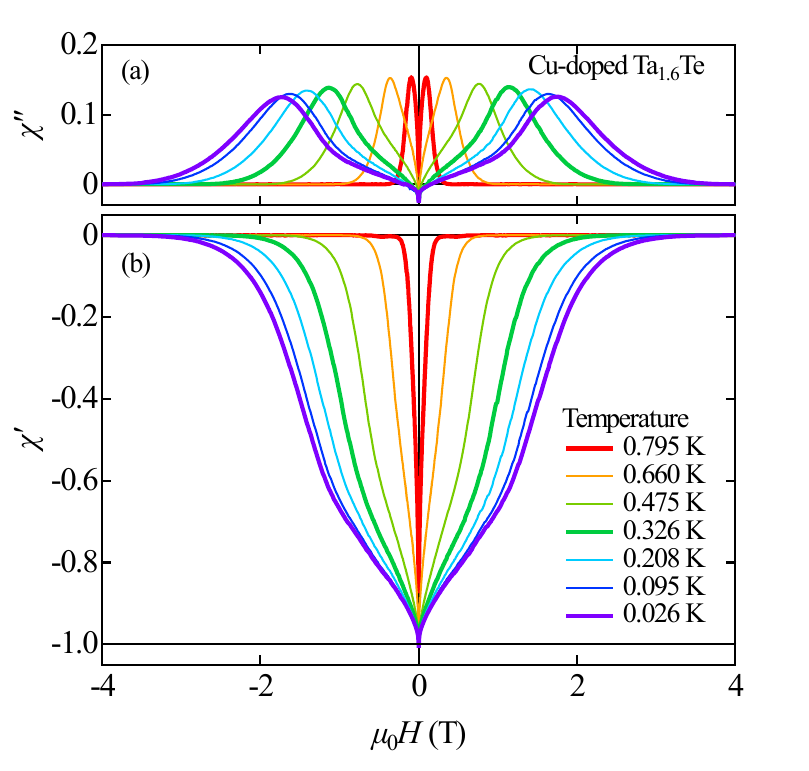}
    \caption{
      Magnetic field dependence of the ac magnetic susceptibility of {\TaTe}.
      Panels (a) and (b) denote the imaginary and real parts, respectively.
      The absolute value of the susceptibility was calibrated against a lead specimen of comparable size and shape as {\TaTe}.
      }
    \label{Fig_ACS}
  \end{figure}

  The field dependence of magnetic properties in aperiodic superconductors has not been reported thus far.
  We therefore measured the ac-susceptibility of {\TaTe}.
  As shown in Fig.~\ref{Fig_ACS}, the real part ($\chi '$) reflects the superconducting shielding signal, while the imaginary part $\chi ''$ represents the energy dispersion due to the flux motion.
  Although $\chi '$ does not prove the Meissner effect, $\chi ' = -1$ indicates that the superconducting current fully excludes flux penetration under an ac field.
  The full shielding signal appears only at zero field, implying that the lower critical field $H_{c1}$ lies below our experimental resolution.
  These results suggest type-{\II} superconductivity in {\TaTe}, consistent with earlier reports for Ta$_{1.6}$Te\cite{Tokumoto2024, Terashima2024}.

  \begin{figure}[t]
    \includegraphics[width=6cm]{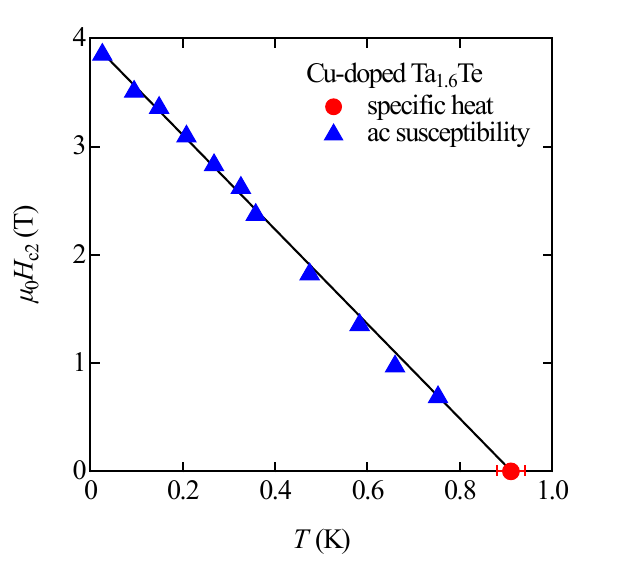}
    \caption{
      Temperature dependence of the upper critical field ($\mu_{0}H_{c2}$) of {\TaTe}, determined from our results.
      The superconducting transition temperature determined from the specific measurement is also shown.
      Error bars for $\mu_{0}H_{c2}$ and temperature stability are comparable to the symbol size.
      }
    \label{Fig_PD}
  \end{figure}

  The shielding signal disappears above the upper critical field ($H_{c2}$).
  We therefore determine $H_{c2}$ as the field where $\chi ' $ vanishes within experimental accuracy.
  The temperature dependence of $H_{c2}$, shown in Fig.~\ref{Fig_PD}, increases linearly with decreasing temperature and extrapolates to $\mu_{0}H_{c2} = 3.95$~T at $T = 0$~K.
  This result agrees well with the resistivity measurement reported for Ta$_{1.6}$Te which gives $\mu_{0}H_{c2} = 4.21$~T\cite{Terashima2024}.


  The BCS theory predicts a universal relation between the superconducting gat and transition temperature as $\Delta_{\rm BCS}(0)/\kB \Tsc \approx 3.53$, independent of material.
  This relation holds for many weak-coupling superconductors, while strong coupling enhances the value\cite{McMillan1968}.
  In contrast, {\TaTe} yields $\Tsc = 0.91$~K and $\Delta(0) / \kB = 1.43$~K, leading to $2\Delta(0) / (\kB \Tsc) = 3.14$, substantially smaller than the of the BCS prediction.
  A similar reduction is seen in the universal relation for the specific-heat jump, where our result $\Delta C / \gamma \Tsc \approx 1.1$ is smaller than the BCS value of 1.43.
  Notably, the prototypical quasicrystal superconductor Al--Zn--Mg also exhibits a small jump at {\Tsc}, consistent with Hubbard-model calculation on the Penrose tiling\cite{Kamiya2018, Takemori2020}.

  We further test whether the small gap $2 \Delta(0) / (\kB \Tsc) = 3.14$ can reproduce the specific heat of {\TaTe}.
  The broken curve in Fig.~\ref{Fig_CiT}(b) represents the calculated specific heat using a scaled gap function with $\Delta(0)/\kB = 1.43$~K.
  The calculation shows a convex feature around 0.5~K due to thermally excited quasiparticles, whereas the experiment exhibits significantly smaller specific heat around this temperature.
  This discrepancy cannot be resolved by varying the fit parameters for electron and phonon contributions in Fig.~\ref{Fig_CiT}(b), and is likely due to suppressed quasiparticle excitation arising from a flattened Bogoliubov peak suggested the Penrose tilling\cite{Takemori2020}, and/or the anomalous drop in $C_{\rm el.} / (T \gamma)$ observed in the normal state.

  Then we deduce the superconducting parameters of {\TaTe}.
  Within thermodynamic and BCS frameworks, the critical field $H_{c}$ and superconducting gap $\Delta(0)$ are related by
  \begin{gather}
    \mu_{0} H_{c}^2 = \frac{3}{2\pi^2} \frac{\gamma}{v} \left( \frac{\Delta(0)}{\kB} \right)^2,
    \label{Eq_HcGap}
  \end{gather}
  where the molar volume $v = 15.1$~cm$^3$mol$^{-1}$ of Ta$_{97}$Te$_{60}$, expected to a similar value for {\TaTe}, is used to convert the per-mol value to per-volume one\cite{Conrad2002}.
  Using Eq.~(\ref{Eq_HcGap}), we evaluate $\mu_{0}H_{c}(0) \approx 6.5$~mT, significantly smaller than $\mu_{0}H_{c2} = 3.95$~T.
  From $H_{c2}$, the coherence length of the Cooper pair $\xi$ is estimated via the Ginzburg–Landau (GL) theory, $\mu_{0}H_{c2} = \phi_{0} / (2 \pi \xi^2)$, yielding $\xi = 9.1$~nm, where $\phi_{0}$ is the magnetic flux quantum.
  These values give a GL parameter $\kappa = 429$ and penetration depth $\lambda = 3.9$~$\mu$m.
  Consequently, $\mu_{0} H_{c1} = 0.14$~mT is available from an approximated formula $\mu_{0} H_{c1} =  \mu_{0}H_{c} (\sqrt{2} \kappa)^{-1} \ln{\kappa}$ for $\kappa \gg 1$.
  The extremely small $H_{c1}$ is consistent with the $\chi'(H)$ curve in Fig.~\ref{Fig_ACS}(b).

  We now turn to the anisotropy in {\TaTe}.
  Although our sample is polycrystalline, the ac-susceptibility results in Fig.~\ref{Fig_ACS} indicate pronounced anisotropy in its superconducting properties.
  Assuming a random grain orientation in the sample, roughly 1/3 of grains are aligned with the out-of-plane field (parallel to the periodic stacking direction), while the remaining 2/3 are aligned with the in-plane field (parallel to the aperiodic plane).
  The slope change of $\chi'$ around $\chi' \approx -0.65$ suggests that approximately 1/3 of grains, under the out-of-plane field, exhibits a smaller $H_{c2}$.
  Since the effective coherence length $\xi$ is defined perpendicular to the magnetic flux, the evaluated $\xi$ primarily reflects the out-of-plane value, whereas the in-plane $\xi$ may be even longer.
  This anisotropy is consistent with expectations from the van der Waals gap and parallels behavior reported in periodic van der Waals compounds\cite{Ribak2020, TaSe2}.
  Studies on single-crystalline samples would provide further insight into the intrinsic nature of this anisotropy.

  Finally, we address the pairing symmetry of the superconductivity in {\TaTe}.
  In periodic systems, a fully gapped states usually implies an isotropic $s$-wave Cooper pairing.
  In contrast, in aperiodic systems, our fully gapped results dose not necessarily indicate $s$-wave pairing, owing to the absence of a well defined $k$-space.
  Theoretical studies of aperiodic superconductivity predict several anomalous features, such as the intrinsic vortex pining, spatially non-uniform order parameter, non-zero momenta, and intrinsic mixing of the $s$- and $p$-wave components\cite{Sakai2017, Sakai2019, Takemori2020, Cao2020, Nagai2022, Fukushima2023, Takemori2024}.
  Moreover, the nature of electronic eigen state under aperiodic potential remains unsettled\cite{Sutherland1986, Tokihiro1988}.
  Thus, capturing both the thermodynamic and microscopic properties of superconductivity, as well as those of the normal electronic state in aperiodic systems, is essential.
  The unexpected drop of the electronic specific heat may represent one such aspect, and further experimental and theoretical studies will be crucial to uncover the true pairing mechanism in quasicrystals.

  In conclusion, we have directly observed the superconducting gap in {\TaTe} by adopting a recently developed technique for specific-heat measurement that eliminates the nuclear contribution.
  The gap symmetry is a fully gaped, but its magnitude and the associated specific-heat jump are both smaller than those expected from BCS theory.
  Quasiparticle excitations are suppressed relative to the calculated full-gap behavior and to predictions from the Hubbard model study on the Penrose tiling.
  The ac susceptibility further reveals an anisotropic critical field, consistent with the quasi-two-dimensional structure of {\TaTe}.
  The upper critical field is much larger than the thermodynamic critical field, pointing to a short coherent length and an extremely small lower critical field.


  The authors would like to thank N. Takemori, N. K. Sato and Y. Mizukami for valuable discussions.
  N.K. is supported by JSPS KAKENHI Grant Numbers 21H01028, 22K03505, and 24K00585.
  Y.T. is supported by JSPS KAKENHI Grant Number 23K04355, Murata Science and Education Foundation and Iketani Science and Technology Foundation.
  K.E. is supported by JST-CREST program (Grant No. JPMJCR22O3; Japan).

\clearpage

\bibliography{TaTe_SC.bib}

\end{document}